\documentclass[11pt]{article}
\newcommand{\Comment}[1]{{}}
\usepackage{amsfonts,amsthm,amsmath,amssymb,slashed}
\usepackage[textwidth = 430 pt, textheight = 630 pt]{geometry}
\usepackage{color}

\Comment{\usepackage{color}
\definecolor{MyDarkBlue}{rgb}{0.15,0.15,0.45}
\usepackage[linktocpage=true]{hyperref}
\hypersetup{
colorlinks=true,
citecolor=MyDarkBlue,
linkcolor=MyDarkBlue,
urlcolor=MyDarkBlue,
pdfauthor={Horatiu Nastase},
pdftitle={The title},
pdfsubject={hep-th}
}

\usepackage[numbers,sort&compress]{natbib}
\usepackage{hypernat}}
\usepackage{graphicx}
\usepackage{cite}

\newcommand\ignore[1]{}
\def\one{{\,\hbox{1\kern-.8mm l}}}

\def\Tr{{\rm Tr\, }}

\def\b{\beta}

\def\d{\partial}

\def\Tr{\mathop{\rm Tr}\nolimits}

\newcommand{\Cset}{{\,\,{{{^{_{\pmb{\mid}}}}\kern-.45em{\mathrm C}}}}}

\newcommand{\be}{\begin{equation}}
\newcommand{\bea}{\begin{eqnarray}}

\newcommand{\ee}{\end{equation}}
\newcommand{\eea}{\end{eqnarray}}

\begin{document}

\renewcommand{\thefootnote}{\fnsymbol{footnote}}

\makeatletter
\@addtoreset{equation}{section}
\makeatother
\renewcommand{\theequation}{\thesection.\arabic{equation}}

\rightline{}
\rightline{}




\begin{center}
{\LARGE \bf{\sc Causal faster than light travel from travel-localized second time coordinate}}
\end{center} 
 \vspace{1truecm}
\thispagestyle{empty} \centerline{
{\large \bf {\sc Horatiu Nastase${}^{a}$}}\footnote{E-mail address: \Comment{\href{mailto:horatiu.nastase@unesp.br}}{\tt horatiu.nastase@unesp.br}}
                                                        }

\vspace{.5cm}


\centerline{{\it ${}^a$Instituto de F\'{i}sica Te\'{o}rica, UNESP-Universidade Estadual Paulista}} 
\centerline{{\it R. Dr. Bento T. Ferraz 271, Bl. II, Sao Paulo 01140-070, SP, Brazil}}

\vspace{1truecm}

\thispagestyle{empty}

\centerline{\sc Abstract}

\vspace{.4truecm}

\begin{center}
\begin{minipage}[c]{380pt}
{\noindent I present a {\em general relativistic} model with a compactified second time coordinate that {\em a priori}
allows for causal, yet faster than light travel in the background of a FLRW geometry, by local modification of a higher 
dimensional background geometry, specifically with respect to the radius of the compactified time coordinate. The modification
can be induced via the fields of the model.
 I show that one cannot convert (as possible in special relativistic models, or simple 
general relativistic models) the super-luminality into closed time-like loops violating causality, due to a novel combination 
of factors, at least for $v_{\rm max}\leq \sqrt{2}$. The physics of the second time is constrained by postulates derived from 
reasonable physical assumptions.
 I comment on the possibility of experimental implications of the model.
}
\end{minipage}
\end{center}

\vspace{.5cm}

\setcounter{page}{0}
\setcounter{tocdepth}{2}

\newpage

\renewcommand{\thefootnote}{\arabic{footnote}}
\setcounter{footnote}{0}

\linespread{1.1}
\parskip 4pt



\section{Introduction}

Since Einstein wrote his special relativistic and general relativistic theories, people have been wondering whether 
it is possible {\em in principle} to travel faster than the speed of light. In special relativity, the answer is a definite NO: 
first, the energy and momentum required to reach the speed of light increases to infinity as $v\rightarrow c$; 
second, there are imaginary quantities involved for $v>c$; and third, assuming the existence of a particle with $v>c$, 
one can construct, by boosting, a closed time-like loop that will violate causality, as I will review later on. 
In 3+1 dimensional general relativity, the answer is a bit more nuanced: there is one notion, first thought of in science 
fiction, of a local disturbance in the gravitational field, a ``warp drive'', a concrete proposal for which was put forward in 
\cite{Alcubierre:1994tu}, that would a allow a spaceship to travel faster than light with respect to an ambient Minkowski 
space, even though locally the ship would always have $v\leq c$. There are several problems with the explicit proposal 
in \cite{Alcubierre:1994tu}, starting with the fact that there is no solution to Einstein's equations, as the metric 
$g_{\mu\nu}$ is just stated; continuing with the fact that it would require ``matter'' (obtained as $T_{\mu\nu}$
being by definition $\frac{1}{8\pi G_N}(R_{\mu\nu}-\frac{1}{2}g_{\mu\nu}R)$) that violates the energy conditions; 
and finally dealing with the same problem, of being able, within the ambient Minkowski space, to create a closed 
time-like loop that violates causality (perhaps there are others objections, but here I focus on these ones). Violations of 
the energy conditions in general relativity can have other effects that lead to superluminal motion and violations of 
causality, from the example of the Morris-Thorne traversable wormhole \cite{Morris:1988cz} (a trip through it 
amounts to superluminal motion), to more recent attempts to create suitable cosmologies, where the focus is on 
{\em excluding } successfully such parameter space from the theory in order to be consistent, 
e.g. \cite{Elder:2013gya}. The null energy condition can be 
violated in the quantum theory, and thus resulting in a minute superluminality, which however cannot be made macroscopic
and/or observable (the earliest paper I know of being \cite{Drummond:1979pp}). 

An unrelated way to {\em a priori} obtain superluminality is to introduce a second time coordinate. For such a space, 
in the absence of any new physics criterion, the violation of causality goes hand in hand with superluminality and is 
easily explained: the second time can change, allowing spatial motion, while the first time is unchanged, and one 
can easily obtain a closed time-like loop, especially if the second time coordinate is considered to be compactified
(since then the second time coordinate {\em is} a loop). One can then impose constraints such that only one time 
is effective, obtaining causal, but sub-luminal physics, as in the long program of two-time physics developed by 
Itzhak Bars and collaborators (some of the relevant references being
 \cite{Bars:1997bz,Bars:1997xb,Bars:1998ui,Bars:2000mz,Bars:2000qm}). 
 
In this paper, however, I want to show that if one considers a few physical conditions, concretized in some principles, 
as well as adding some somewhat obvious field to the theory, consistent with both the principles and observations, 
we find a theory that has the potential to have {\em causal} superluminality. Physics still has effectively one time only, 
through a constraint arising from the principles, and as observed in everyday life, but the presence of the second time can 
lead to superluminality (from the point of a fictitious Minkowskian background that we can use for observations), provided
one can find a way to excite a field of the theory. I will show that it is now impossible to construct a closed timelike loop
using the superluminality. A necessary ingredient is the existence of a spontaneous breaking of the (fictitious) Lorentz 
invariance consistent with the presence of the background FLRW cosmology, as previewed in \cite{Nastase:2006na}.

The paper is organized as folllows. In section 2 I describe the model with compactified time. In section 3 I introduce some physical 
principles, and resulting from them, I quantify the {\em apparent} Lorentz violation. In section 4, I describe the fact that this violation can be
nevertheless causal. In section 5, I present some observations on fields and quantization in the presence of the second time, and some 
ideas about observational consequences, and in section 6 I conclude.

\section{Compactified second time and model}

The basic idea of the paper is that we are in a general relativistic theory, and
 there is a small, compactified, second time-like coordinate, so that in a certain gauge 
for  general coordinate invariance (meaning that we can put $g_{0i}=0$), we have 
\be
ds^2=-|g_{00}|dt^2-|g_{0'0'}|dt'^2+g_{ij}dx^idx^j\;,\label{ansatz}
\ee
but where $t'$ is periodic with a small period (radius). In the following, I will always denote this second time with $t'$, or 
$x^{0'}$. Additionally, I will use theorist's units and put the speed of light in Minkowski space (at least 
with respect to the usual time $t$) to one, $c=1$, and denote by $x^i$ the 3  large spatial coordinates that we observe
(there could be more compact spatial coordinates, like in string theory for instance, but we will ignore those in this paper). 

I will also assume that the periodicity of $t'$ is $2\pi R_0$, meaning that $t'$ is dimensionful like $t$,
and then $|g_{0'0'}|=R^2/R_0^2$ measures the radius in units
of $R_0$. Moreover, I will assume that $|g_{0'0'}|=\epsilon\ll 1$. We will specify the meaning of $R_0$, and the 
splitting of the periodicity into $R_0$ and $\sqrt{\epsilon}$, in the 
next section.

The implication of the condition $|g_{0'0'}|\ll 1$ is that under normal circumstances, the path of light, $ds^2=0$, 
is mostly unaffected by the extra term. But if we can increase $|g_{0'0'}|$, thought of as a scalar field that can condense, 
just like in any regular Kaluza-Klein model, we can have an observable effect. The condensation of $\phi$ can be driven 
by some other matter field, which we must define. 

If gravity is coupled to matter, the natural candidate that can interact with the metric and condense $\phi$ is an 
antisymmetric tensor field strength. Since the background spacetime is actually the cosmological FLRW metric for 
the expanding Universe, such a metric has a preferred Lorentz frame with respect to the tangent Minkowski space, valid in a
small neighbourhood (since Lorentz invariance is not a symmetry of the FLRW metric). This defines the cosmic time $t$ 
in the metric, and an apparent ``spontaneous breaking of Lorentz invariance'': to be more precise, since we are in 
general relativity, we only have {\em local } Lorentz invariance, which is always preserved, but global Lorentz invariance 
is broken by the choice of time in the FLRW metric. This is why we can detect the Earth's motion with respect to the system 
of the CMBR, which defines the cosmic time. That means that in the FLRW metric, we can have a special vector with only 
nonzero component $A_0$ (where $x^0$ here is the cosmic time), or equivalently a special 3-index antisymmetric
tensor with spatial components only, $F_{ijk}$, with a single independent component, $F_{123}$, as considered in 
 \cite{Nastase:2006na}. 

Since we want to have an extra time direction, we can have a possible 4-index antisymmetric tensor, with only nonzero 
independent component $F_{(4)0'123}$. Since this should be the field that condenses, $\phi$, we want that there is a 
field strength $F_{(4)MNPQ}$ interacting with gravity, 
where we denote by $M,N,...=0,0',1,2,3$ the (3+2)-dimensional space, and by 
$\mu,\nu,...=0,1,2,3$ the (3+1)-dimensional space. We want that normally, $c=1$ should be the maximum speed, 
so we must have $|g_{0'0'}|=\epsilon\ll 1$ normally, and then $F_{(4)0'123}\simeq 0$ as well, but when $F_{(4)MNPQ}\neq
0$, we want $|g_{0'0'}|\neq 0 $ (i.e., macroscopic, or of order one) also. 

\subsection{Model and its equations of motion}

The action for the model is then 
\bea
S&=&\frac{M^3_{\rm Pl,5}}{2}\int d^5x \sqrt{+g^{(5)}}R^{(5)}+\frac{M^3}{2}\int d^5x\sqrt{+g^{(5)}}f
\left|\frac{1}{4!}F_{(4)MNPQ}F_{(4)}^{MNPQ}-m^2\right|\cr
&&+\int \sqrt{+g^{(5)}} {\cal L}_{\rm matter}.\label{action}
\eea
Here $M_{\rm Pl,5}$ is a 5-dimensional Planck scale, $M$ is an independent quantity, $g^{(5)}>0$ since we have a
3+2 dimensional signature, ${\cal L}_{\rm matter}$ is the Lagrangian for matter other than $A_{(3)MNP}$ and of $m$, 
$F_{(4)MNPQ}=4\d_{[M}A_{NPQ]}$ (total antisymmetrization with strength one), so that $F_{(4)}=dA_{(3)}$ in form 
language, and $m$ is a quantum VEV of matter fields, about which we will say more later on. 
Moreover, $f$ is a scalar function that will be defined shortly
in several cases, and the absolute value in this term is present only in the $f=1$ case. 

For economy of the model, we can choose $M=M_{\rm Pl, 5}$, which avoids annoying factors of 
$M^3/M_{\rm Pl,5}^3$, which can otherwise be reabsorbed in the definitions of $A_{(3)MNP}$ and $m$.

Consider its equations of motion, under the dimensional reduction ansatz for the metric with respect to $t'$,
\be
\d_{0'}g_{MN}=0.
\ee                                                                                                                 

The most relevant component is the Einstein equation of $g_{0'0'}$. 

On the ansatz, we calculate first
\bea
R_{0'0'}&=&\d_M {\Gamma^M}_{0'0'}+{\Gamma^M}_{NM}{\Gamma^N}_{0'0'}-{\Gamma^M}_{N0'}{\Gamma^N}_{M0'}\cr
{\Gamma^M}_{0'0'}&=& -\frac{1}{2}g^{MN}\d_N g_{0'0'}\cr
{\Gamma^M}_{0'N}&=& \frac{1}{2}g^{MP}(\d_N g_{P0'}-\d_P g_{N0'})\Rightarrow\cr
{\Gamma^M}_{0'M}&=&0\;,
\eea
obtaining 
\bea
R_{0'0'}&=& -\frac{1}{2}\d^M\d_Mg_{0'0'}-\frac{1}{2}(\d^\mu g_{0'0'}){\Gamma^M}_{\mu M}-\frac{1}{4}(\d^Q g_{P0'}\cr
&&-g^{NQ}\d_P g_{N0'})
(\d^P g_{Q0'}-g^{MP}\d_Q g_{M0'}).
\eea
Next, we use the identities
\bea
{\Gamma^M}_{\mu M}&=& {\Gamma^\nu}_{\mu\nu}+\frac{1}{2}g^{0'0'}(\d_\mu g_{0'0'})\Rightarrow \cr
D_\mu^{(5)} \d^\mu &=&\d_M \d^M +{\Gamma^M}_{\mu M}\d^\mu =D_\mu^{(4)} \d^\mu
+\frac{1}{2}g^{0'0'}(\d_\mu g_{0'0'})\d^\mu\;,
\eea
to finally write $R_{0'0'}$ in the form
\be
R_{0'0'}=-\frac{1}{2}D_\mu^{(4)} \d^\mu g_{0'0'}+\frac{1}{4}g^{0'0'}(\d_\mu g_{0'0'})(\d^\mu g_{0'0'}).\label{R00}
\ee

The first observation then is that the Eistein equations in vacuum (for no matter contribution, and 
$F_{(4)MNPQ}\simeq 0\simeq m$), in particular the $0'0'$ component,
\be
R_{0'0'}-\frac{1}{2}g_{0'0'}R=0\;,
\ee
is solved by 
\be
g_{0'0'}=\epsilon\rightarrow 0\;,
\ee
since in that case we also obtain 
\be
R_{0'0'}\rightarrow 0.
\ee
Moreover, the contribution of $g_{0'0'}$ to the $R_{\mu\nu}$ components, $\mu,\nu=0,1,2,3$, is also negligible, and 
we are left with the usual 3+1 dimensional Einstein's equations in vacuum.

This is the solution that we expect in a normal situation: then the radius of the extra time dimension is both very small,
and its effect on physics is also infinitesimal, 
and we can effectively ignore it. 

Next, we move on to the case of nontrivial matter, in particular nonzero $F_{(4)MNPQ}$ and $m$.
The Einstein equations are 
\bea
&&R_{MN}-\frac{1}{2}g_{MN}R\cr
&=&f\left[\frac{4}{4!}F_{(4)MPQR}{F_{(4)N}}^{PQR}-\frac{1}{2}g_{MN}
\left(\frac{1}{4!}F_{(4)PQRS}^2-m^2\right)\right]sgn\left(\frac{1}{4!}F_{(4)}^2-m^2\right)\;,\cr
&&\label{Einstein}
\eea
and the 3-form equations of motion, for $A_{(3)MNP}$, are 
\be
\d_M (\sqrt{g_{(5)}} F^{(4)MNPQ})=0\;,
\ee
which need to be supplemented with the Bianchi identities
\be
\d_{[M}F_{(4)NPQR]}=0.
\ee

Consider {\em static } solutions, $\d^0({\rm anything})=0$.
The equations of motion of the 3-form are, expanding in components $NPQ=ijk; 0ij;00'i$ and $0'ij$, 
\bea
&&\d_{0'}(\sqrt{g_{(5)}}F^{(4)0'123})=0\cr
&&\d_k(\sqrt{g_{(5)}} F^{(4)k0ij})+\d_{0'}(\sqrt{g_{(5)}}F^{(4)0'0ij})=0\cr
&&\d_j (\sqrt{g_{(5)}} F^{(4)j00'i})=0\cr
&&\d_k (\sqrt{g_{(5)}} F^{(4)k0'ij})=0.\label{F4eom}
\eea

All in all, we obtain that $\sqrt{g_{(5)}}
F^{(4)0'ijk}$ is constant in space ($x^i$) and in $0'$, as well as in time $0$, so is a constant, 
\be
\sqrt{g_{(5)}}F^{(4)0'ijk}={\rm constant}\equiv m\;,\label{4form}
\ee
as we needed. 

We can define the Poincar\'{e} dual to $A_{(3)MNP}$, which is a scalar $\phi$, by
\be
F^{(4)MNPQ}=\frac{1}{\sqrt{g_{(5)}}}\epsilon^{MNPQR}\d_R\phi.
\ee
In particular then, we obtain 
\be
F^{(4)0'123}=\frac{1}{\sqrt{g_{(5)}}}\epsilon^{0'1230}\d_0\phi.
\ee
Here we have used the standard choice for the Levi-Civitta tensors for the object with indices up to be 
pure number, $\epsilon^{0'123}=\epsilon^{0'1230}=+1$, and when lowering the indices we obtain factors of the metric. 
But given (\ref{4form}), we obtain 
\be
\d_0 \phi=m\Rightarrow \phi=mt+\phi_0.
\ee

Then 
\be
F_{(4)MNPQ}F^{(4)MNPQ}=4!g^{00}\d_0 \phi \d_0\phi=4!(\d_M\phi)^2=4!m^2 g^{00}\;,
\ee
which puts to zero the $F_{(4)}$ term in the action (\ref{action}).

Here we should remember however that we cannot simply trade $A_{(3)MNP}$ by its Poincar\'{e} dual $\phi$, since 
their energy-momentum tensors are different. We can only do so {\em after} considering Einstein's equations. Indeed, 
the energy-momentum tensor term
\be
\frac{\delta F^2}{\delta g^{0'0'}}=4F_{0'ijk}{F_{0'}}^{ijk}=4 \frac{F_{MNPQ}F^{MNPQ}}{4 g^{0'0'}}=4!m^2
g_{0'0'}g^{00}=g_{0'0'}(F_{MNPQ}F^{MNPQ})
\ee
differs from the one of the dual field, 
\be
\d_{0'}\phi \d_{0'}\phi\simeq 0.
\ee
This is as it should be, since the variation with respect to the metric is when keeping the {\em fundamental field }
fixed, not the constants in its classical value. Thus in the action we must consider $A_{(3)MNP}$ and not simply a 
scalar, as we would be tempted to do.

The equations of motion (\ref{F4eom}) imply also that $F_{(4)00'ij}$ is constant in space. Consider next the Bianchi 
identities, which reduce to the single equation
\be
\d_{[M}F_{(4)NPQR]}=0\;,
\ee
or explicitly 
\be
\d_{0'}F_{(4)0123}-\d_0 F_{(4)0'123}+\d_1 F_{00'23}+\d_2F_{00'31}+\d_3 F_{00'12}=0.
\ee
But by the dimensional reduction assumption $\d_{0'}F_{(4)0123}=0$, and by the static assumption 
$\d_0 F_{(4)0'123}=0$, so 
the Bianchi identity is consistent with the solution of $F_{(4)00'ij}$ being constant in time, as well as in space. 
We can therefore {\em choose} to put it to zero, as there is no 
equation of motion that requires it to be nonzero.

In conclusion, as a solution we can choose $F_{(4)0'123}\neq 0$ and $F_{(4)00'ij}\simeq 0$. 
The solution must moreover minimize the potential, which is done by putting to zero the absolute value in (\ref{action}). 
This was implicit in (\ref{4form}), where we denoted the constant by $m$. 
We then  have the solution 
\be
F^{(4)0'ijk}=m \frac{\epsilon^{0'ijk}}{\sqrt{g_{(5)}}}\;,\label{F4m}
\ee
which means
\be
F_{(4)0'ijk}=m \frac{\epsilon^{0'ijk}}{\sqrt{g_{(5)}}} g_{0'0'}g_{(3)}.
\ee

Next, consider the Einstein equations (\ref{Einstein}). Multiplying them with  $g^{MN}$, we get
\be
-\frac{3}{2}R=\frac{4}{4!}fF^2-\frac{5}{2}f\left(\frac{F^2}{4!}-m^2\right)\;,
\ee
with the last term chosen to vanish, so $R=- F^2/9\sim m^2$. Here $R=g^{0'0'}R_{0'0'}+g^{00}R_{00}+g^{ij}R_{ij}$.

Again assuming the potential to vanish as before (yet the $sgn$ be $+$ always), the $0'0'$ component becomes
\be
R_{0'0'}-\frac{1}{2}g_{0'0'}R=\frac{f}{3!}F_{(4)0'PQR}{F_{(4)0'}}^{PQR}\;,
\ee
so it looks like we must have a nonzero $g_{0'0'}$ now. Except that {\em on the solution}, 
\be
4 F_{(4)0'ijk}{F_{(4)0'}}^{ijk}=g_{0'0'}(F_{(4)MNPQ}F^{(4)MNPQ})\;,
\ee
so in effect, the whole equation is proportional to $g_{0'0'}$, so we haven't proven anything yet. We must look further 
to do so.

From the $00$ and $ij$ components of the Einstein equations, 
\bea
&& R_{00}-\frac{1}{2}g_{00}R=0\cr
&& R_{ij}-\frac{1}{2}g_{ij}R=4F_{iMNP}{F_j}^{MNP}\;,
\eea
by multiplying them by $g^{00}$ and $g^{ij}$, respectively, and subtracting from the equation $R=- F^2/9\propto m^2$,
we obtain
\be
g^{0'0'}R_{0'0'}\propto F^2\propto m^2.
\ee

But then, because of (\ref{R00}) it follows that, generically, 
\be
g^{0'0'}\d_{\mu}g_{0'0'}\propto m.
\ee

In turn, that means that $g_{0'0'}$ varies as $\sim m x^\mu$, so it must be nonvanishing. Therefore we have 
proven what we set out to do, that in the presence of nonzero $m$ and 
correspondingly nonzero $F_{(4)0'123}$, we have nonzero $g_{0'0'}$. 

The last thing to understand is the meaning of the scalar function $f$ in the action for $A_{(3)MNP}$. It was introduced 
so that the we can have a simple reason to minimize the potential and obtain (\ref{F4m}) as an equation of motion. 
Of course, one possibility is to keep $f=1$ as well as the absolute value, in which case minimization gives the absolute 
value vanishing, leading to (\ref{F4m}). But that is  perhaps not too satisfactory, since there is no associated field that 
can give it as equation of motion. Therefore, we have 3 possibilities:

\begin{itemize}

\item The possibility stated, that $f=1$ and absolute value included. 

\item We can impose (\ref{F4m}) as a constraint with a Lagrange multiplier, so $f=\lambda$ is a Lagrange multiplier 
(non-dynamical field). Redefining both $A_{(3)MNP}$ and $m$, we can then rewrite the term as 
\be
-\int d^5x\sqrt{g_{(5)}} \lambda [F_{(4)MNPQ}{F_{(4)}}^{MNPQ}-m^4]\;,
\ee
where now $A_{(3)}$ has dimension 1, as does $\lambda$, and $m^4$ comes as before from matter.

\item We can make $\lambda$ dynamical, giving it a kinetic term, $-M \int \sqrt{+g_{(5)}}\frac{1}{2}(\d_M\lambda)^2$. 
But if this is a dilaton-like scalar, like it is common to have in string theory, the kinetic term is not canonical, and by 
making it canonical we obtain an exponential function $f$, specifically we obtain 
\be
-M\int d^5x\sqrt{g_{(5)}}\left[\frac{1}{2}(\d_M\lambda)^2+e^{-\lambda/M'}
 [F_{(4)MNPQ}{F_{(4)}}^{MNPQ}-m^4]\right].
\ee

\end{itemize}

\section{Physical principles and quantifying the apparent Lorentz violation}

Now that we define a model with two times, that has a chance of causing an apparent Lorentz violation, we have to understand what does the model mean 
in the context of general relativistic physics. Since this is an extension of usual general relativity that involves a new element, we have to extend the fundamental 
principles of general relativity to include this new information. We can understand Einstein's general relativity as being based on the 
two principles of: 1. Physics is invariant under general coordinate transformations, and 2. The equivalence principle. These, in turn, are based on the two 
physical assumptions, that A. "Gravity is geometry", and B. "Matter sources gravity", at which Einstein arrived through an analysis of consistency of 
special relativity with gravity. 

Similarly then, we should now search for physical assumptions about the presence and meaning of two times in physics, and turn them into physical principles. 
Unlike the case of general relativity above, there is no actual {\em need} to have such principles (there is no theoretical conflict to resolve, like the one between 
special relativity and gravity), but I simply propose that this is a possibility that has to be taken into account: if we can think about having more than 3 spatial 
coordinates, we can also think about having more than one time coordinate.

I will states these principles, or postulates, and then comment on which physical assumptions are behind them, and how they are arrived at. 

{\bf Postulate 1. Light moves on null geodesics.}

We have to still impose that light moves on $ds^2=0$, in order to be able to still distinguish between space-like and time-like intervals, 
and the boundary between them to be the motion of light. We also need  still that the motion of light is a geodesic motion, in order not to interfere with 
the usual general relativity principles. We can summarize this by saying that the physical assumption is "light distinguishes between space and time".

{\bf Postulate 2. There is a single "affine parameter" on the motion of a particle}.

This affine parameter is thought of as the "proper time" on the worldline of a particle. In other words, having two times doesn't mean that the particle 
has a worldsheet in the two times; it must still have a worldline. This fact 
is necessary for sanity, since we know that in our case we don't experience anything other than linear time. In other words, the physical assumption is
"not just people, but any particle must only experience linear time".

{\bf Postulate 3. The ratio of normal time to second time is a universal constant.}

This third postulate is in some sense derived from the second one, though only in part. We must quantify the effects  of  $g_{0'0'}=\epsilon\rightarrow 0$ as being 
small: the effect of this fact {\em now} (in the vacuum) must be infinitesimal, so unobservable, but nonzero for them to become useful if $g_{0'0'}$ is changed. 
Small $g_{0'0'}$ means that there is a small radius for the extra time; but if the ratio of the two times is not fixed in terms of its effect inside the single affine 
parameter of postulate 2, a small radius could still have large effects. Then we must have
\be
\frac{dt'}{dt}=\frac{dx^{0'}}{dx^0}={\rm constant}.
\ee
Since $c=1$ in theorist's units (due to the constancy of the speed of light at $g_{0'0'}=0$), the simplest choice,
which amounts to another choice of units, besides the one implied by $c=1$, is to set this constant also to 1, 
\be
\frac{dt'}{dt}=1.
\ee
But this only has meaning if we define {\em units} for $t'$, just like there are units (the meter, if $c=1$) for $t$. We said that the periodicity of $t'$ is $2\pi R$, which 
has implicit the notion that the units for $t'$ are also defined {\em in} meters, and the meaning of the value $R$ is that we have traded the ratio $dt'/dt$ for 
the periodicity of $t'$ (which otherwise could have been chosen as $2\pi$). The result is that $g_{0'0'}$ now is a dimensionless variable that makes sense: the physical 
periodicity of the compact coordinate is $2\pi R \sqrt{g_{0'0'}}$, and is a value independent of the value of the constant $dt'/dt$. 
Finally, the physical assumption behind this third principle can be summarized as "the compact second time under normal circumstances has negligible effects".

{\bf Postulate 4. The universal constant is valid only in a preferred frame defining $t$.}

This fourth postulate is also partly derived from the previous two, but contains new information. 
As I mentioned, the existence of the cosmic frame defined by the FLRW background metric means that there is a preferred time, in which the CMBR is at rest, and it is 
this frame that is referred in the postulate. 
The existence of this preferred time
 doesn't contradict general relativity, since we are talking about a {\em global } Lorentz violation, which is anyhow meaningless in general relativity. 

Why do we need this fourth postulate? Because only if $dx'^0=dx^0$ in the preferred frame only, is $x'^0$ an {\em independent} coordinate. Otherwise 
it amounts to just a (trivial) rescaling of the usual time coordinate. So the physical assumption involved here is "the second time is an independent coordinate".

Note that the equality in the preferred frame only naively sounds as if it implies a breaking of general coordinate invariance, but in fact it is exactly the opposite: it is the only 
way to preserve it.
Indeed, now in the special frame we have the proper time
\be
-d\tau^2\equiv ds^2=-dx_0^2-g_{0'0'}dx_{0'}^2=-dx_0^2(1+g_{0'0'}).\label{propertime}
\ee

But away from this frame, we transform $x_0$ and the direction of motion $x_1$, but not the second time $x_{0'}$! Then the invariance means that we have
\be
-dx_0^2=-d\tilde x_0^2+d\tilde x_1^2\;,
\ee
but also $g_{0'0'}dx_{0'}^2$ stays invariant, since $dx_{0'}$ equals $dx_0$, not $d\tilde x_0$, meaning that in total $d\tau^2=-ds^2$ 
stays invariant to the reference frame change, as is natural to be! General coordinate invariance is thus maintained only in this case. 

It might seem strange that we need the existence of the special frame, which in turn relies on the FLRW expanding Universe, for a fundamental definition of the second 
time. But perhaps one could argue also in reverse, that the second time necessitates and implies an expanding FLRW Universe, though I will not examine further this 
possibility, and it is left for further work.

I have presented here an "axiomatic system" made up of 4 postulates, but probably this can be further reduced. I just presented it this way, since I found it easier to 
present its "derivation" from physics.

\subsection{Apparent Lorentz violation}

I now turn to understanding its implications for an apparent Lorentz violation. Since $dt'/dt=1$ and $ds^2=0$ for light, we can write for light
\be
ds^2=d\vec{x}^2-dt^2-g_{0'0'}dt'^2=0.
\ee
But then the the maximum {\em local} speed is the true speed of light, defined by this $ds^2=0$, implying
\be
v^2_{\rm max}=\frac{d\vec{x}^2}{dt^2}=1+g_{0'0'}\frac{dt'^2}{dt^2}=1+g_{0'0'}.
\ee
Here $c=1$ is the speed of light in terms of the approximately Minkowski space away from the disturbance with nonzero $g_{0'0'}$ ("at infinity", though not quite, 
since then we must remember that we have an FLRW background at very large scales), so is the "speed of light from the point of the faraway observer". 
Then for this faraway observer, we have an apparent maximum Lorentz violation quantified by 
\be
v^2_{\rm max}-1=g_{0'0'}\;,
\ee
though we can only state that with precision in the preferred frame. 

More precisely, we find that $v^2_{\rm max}-1$ decreases away from the preferred frame. 

In a moving frame, $d\tilde t=\gamma dt$, which can be {\em reinterpreted} as 
$\tilde g_{0'0'}=g_{0'0'}/\gamma^2$, but really the metric is not changed. The {\em physical statement} is that for light, defining the maximum speed,
\be
ds^2=0=-d\tilde x_0^2+d\tilde x_1^2-\frac{d\tilde x_0^2}{\gamma^2}g_{0'0'}
\Rightarrow \tilde v_{\rm max}\equiv \frac{d\tilde x_1}{d\tilde x_0}\;,
\ee
which means that the maximum local speed satisfies
\be
\tilde v_{\rm max}^2-1=\frac{g_{0'0'}}{\gamma^2}=\frac{v^2_{\rm max}-1}{\gamma^2}.
\ee

Thus away from the preferred reference frame, the apparent Lorentz violation is smaller.

\section{Causal Lorentz violation}

We want to show that this result means that we can still preserve causality, even though we move at $v_{\rm max}>1$. To understand how is that possible, 
and what are the crucial new elements that make this result possible, we build up the answer by investigating other cases, and seeing what was missing in each. 

{\bf Example 1.  Galilean space (special vector)}. 

The first example to try is obviously Galilean physics, what one used before Einstein, which allowed for faster than $c$ travel that doesn't 
contradict causality. The way it does that is by selecting a preferred time direction, one that is {\em not changed} under Lorentz transformations, 
thus breaking Lorentz invariance. Therefore under Lorentz transformations, we have $dt'^2=dt^2$ and, since 
\be
ds^2=g_{\mu\nu}dx^\mu dx^\nu\;,
\ee
we have an invariant tensor $g_{\mu\nu}\equiv B_{\mu\nu}={\rm diag}(1,0,0,0)$ or, thinking of $B_{\mu\nu}$ as the product of two vectors, $B_\mu C_\nu$, a constant 
vector $B_\mu=(1,0,0,0)$ selecting the time direction.

The problem is, of course, that there is no such vector, since we don't live in Galilean space, but in one that is locally Minkowski.

{\bf Example 2.  Localized curved space}

The next example is of a more general type, and is already in general relativity, i.e., in curved space. Consider a {\em localized } distorsion of a mostly 
Minkowskian spacetime, by curving the spacetime around a desired superluminal trajectory $x^\mu(t)$, such that the {\em local} speed 
of light in some part of the distorsion corresponds to a superluminal velocity from the point of view of the coordinates outside the distorsion. 
This is the idea behind Alcubierre's "warp drive" \cite{Alcubierre:1994tu} mentioned in the introduction, and one is supposed to "surf the wave" of the gravitational distorsion. 

Of course, even assuming that the ``warp drive'' works to send a super-luminal ship (which is not clear it does), 
Alcubierre has not solved anything, since writing an arbitrary metric, and calculating the resulting energy-momentum tensor 
from Einstein's equations doesn't mean solving anything. Moreover, the resulting $T_{\mu\nu}$ contains regions of negative energy density. As I said, that is 
true of all known ways to obtain super-luminal motion (like the Morris-Thorne traversable wormhole, for instance), and anyway 
that could be solved {\em in principle} by something similar to the construction with two times, that would 
perhaps generate effectively negative energy density, though we would 
still need to show that we could build something like the warp drive by using such $T_{\mu\nu}$. 

However, here we assume that such a construction is possible, and we are interested to see whether it is possible for it to avoid causality violations. 

In order to see that, we first must understand how is it possible to construct causality violation from superluminal motion in Minkowski space. 
The short answer is that we must boost the system (using subluminal velocity) and then use the superluminal motion backwards in the boosted system. 
First, in an $(x,t)$ diagram, 
boosting the coordinate system corresponds to tilting the orthogonal axis $x$ and $t$ inside, to $(x',t')$ that have a smaller angle between them. 
Indeed, a boost is a coordinate transformation 
\bea
x'&=&x\cosh \b -t\sinh \b \cr
t'&=& -x\sinh \b +t\cosh \b\;,
\eea
which gives $x'^2-t'^2=x^2-t^2$. In the $(x,t)$ plane, the $x'$  line is $t'=0$, which is
\be
\frac{x}{t}=\frac{\cosh \b}{\sinh\b}>1\;,
\ee
and the $t'$ line is $x'=0$, which is 
\be
\frac{x}{t}=\tanh \b \in [0,1]
\ee

\begin{figure}[h!]
\begin{center}
\includegraphics[width=130mm]{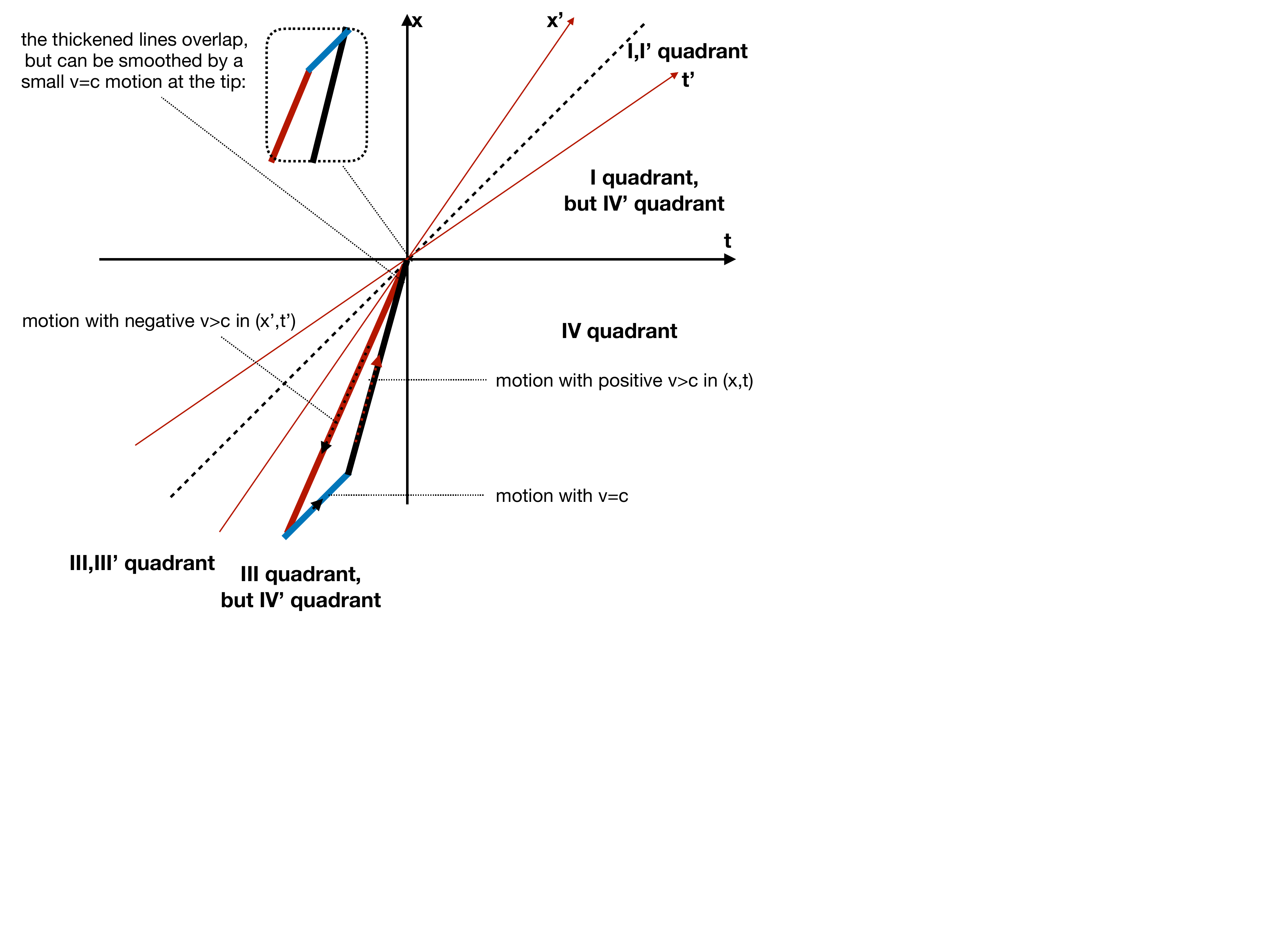}
\end{center}
\caption{The closed time-like loop constructed from velocities with $v>c$ in both $(x,t)$ and $(x',t')$ coordinate systems.}
\label{fig:CTLoop}
\end{figure}

To construct a closed "time-like" (motion) 
loop in this diagram, consider the following set-up. Consider first a motion in the negative $x'$ direction (but positive 
$t'$ direction), but with sufficiently large $v>c$, such that, while it is in the IV quadrant from the point of view of $(x',t')$, because these are not 
rectangular quadrants in the $(x,t)$ picture, they become in the III quadrant (negative $x$, negative $t$ directions) in the $(x,t)$ picture. Then a short 
motion with $v=c$ in the positive $x$, positive $t$ direction, followed by a motion with sufficiently large 
$v>c$  in $(x,t)$ system that takes us from the point in the III quadrant to the origin (so motion in the positive $x$, positive $t$ direction). 
The loop is described in Fig.\ref{fig:CTLoop}.

Physically, this corresponds to moving with $v>c$ backwards in the system moving forwards ($x/t>0$), then at (close to) the speed of light forwards, 
then at $v>c$ forwards. 

Now, in the localized curved space, we assume that the space is curved around the trajectory, so the $v>c$ lines of the loop become "thick", because 
of the gravitational field which should somehow be generated around the trajectory. But obviously, we should not have a situation where the 
two thicknesses touch, since then we cannot generate one field due to the presence of the other. 

That seems to mean that the loop becomes impossible, since at the origin the two thicknesses could touch, but that would be easily fixed, 
by just adding a small line of (almost) speed of light $v=c$ motion separating the two lines. So the only way for this loop to be illegal would be if it 
would not be possible to generate the $v>c$ line in $(x',t')$ that moves backwards in $t$, since it would  "pass over" the $v>c$ line in 
the $(x,t)$ system in order to get from the IV quadrant to the III quadrant. But it is hard to see why this would be the case within normal general 
relativity. After all, the background space is Minkowski, and boosting should produce no change in the ability to construct the "warp drive" as defined, 
so there is no a priori reason for it. 

{\bf Example 3. FLRW space}

In the case of FLRW space, as the space is not asymptotically Minkowski, it is not surprising that "the speed of light is greater than c", in the sense that as light passes by, the space
expands, so in time $t$, light passes more than $x=t$, depending on $a(t)$. For instance, we have $x=pt=2t$ for 
$a(t)=t^n$.
Of course, locally, the speed of light is still $c$, and any velocity is bounded by it, since we have 
local Lorentz invariance.

More importantly, in this case, {\em there is no }{\bf global} {\em Lorentz invariance}, on the contrary, there is a preferred time coordinate $t$, for comoving coordinates. The 
corresponding system of coordinates, called sometimes the CMBR system,  can be experimentally detected: we know precisely the speed of motion of the Earth and Sun 
with respect to this system of coordinates, i.e., with respect to the CMBR. 

Formally, this {\bf global} Lorentz violation is defined, as in the case of Galilean space, by the presence of a constant Lorentz vector $V_\mu=(1,0,0,0)$ that defines the time
coordinate, leading to the FLRW metric (assuming $k=0$, i.e., flat spatial Universe)
\be
ds^2=-dt^2+a^2(t)(dr^2+r^2d\Omega_2^2).
\ee

Of course, this Lorentz violation doesn't help to travel with $v>c$ along some trajectory, since we can only travel at most at $v=p$, but moreover 
the space itself recedes away from us at the same rate, so that doesn't help at all.

 {\bf Example 4 (solution). FLRW space + special vector + curved space}

Finally we arrive at the solution that I propose. We consider an FLRW spacetime solution in the theory with two times described before, with $g_{0'0'}\simeq 0$ in the 
background. Moreover, we assume that we can 
construct a localized curved space, but now meaning nonzero (macroscopic) $g_{0'0'}$  along a particular trajectory $x^\mu(t)$. 
As we saw, this means a nonzero (approximately constant over the nontrivial region needed along the trajectory $x^\mu(t)$) tensor $F_{(4)0'123}$, which is consistent 
with the symmetries of the FLRW metric, and is equivalent (by dualization) with a constant vector $C_0$, i.e., $C_\mu=(1,0,0,0)$ in the $(0123)$ space. 

In some sense then, this proposed solution is the sum of the 3 previous examples. As such, we can think from the point of view of each of the 3 previous 
cases, and see how we fixed the problems with them:

\begin{itemize}

\item From the point of view of Galilean space, we have the now 
constant $C_0$, {\em but only along the trajectory $x^\mu(t)$}, thus not contradicting relativity. The existence of the FLRW metric is essential for its construction (provides
the symmetries needed for constructing $C_0$)

\item From the point of view of the localized curved space, now we have also the constant vector, which means that the two boosted frames {\em are not equivalent}, and 
the picture we had, of creating a $v>c$ in the boosted frame, is not possible. More precisely, we saw that by boosting with a parameter $\gamma$, the 
Lorentz violation in the new system, $v^2_{\rm max}-1$, decreases by the same factor $\gamma$.

\item From the point of view of the FLRW space, now we have the localized curved space along the desired trajectory, so we can travel faster than $c$ along it, and 
$C_0$ along the trajectory means the same thing.

\end{itemize}

We now turn to the constraints imposed in order to avoid the existence of the loop at example 2. 

The motion with $v>c$ in $(x,t)$ in the loop occurs for (in the limiting case, of the maximum velocity) 
$x/t=v_{\rm max}$. For the system to be causal, we need to not be able to create the line that goes back in time before 
the beginning of the $x/t=v_{\rm max}$ line, even in the extremal case when we consider motion with $\tilde v_{\rm max}$
on the negative $x'$ axis. That is, we want that the line $x'/t'=-\tilde v_{\rm max}$ is either steeper than the $x/t=v_{\rm 
max}$ line, or in the fourth quadrant. Then we need that ($v=\tanh \b$ is the velocity of the boost of the coordinate system)
\bea
\frac{x'}{t'}&=&\frac{x\cosh \b -t \sinh \b}{-x\sinh \b +t \cosh \b}=-\tilde v_{\rm max}\Rightarrow\cr
\frac{x}{t}&&=\frac{v-\tilde v_{\rm max}}{1-v\tilde v_{ max}}
\eea
obeys
\bea
\frac{x}{t}=\frac{v-\tilde v_{\rm max}}{1-v\tilde v_{ max}}\geq v_{\rm max}\;\;\; {\rm OR}\;\;\;\; \leq 0.
\eea

In the second case ($\leq 0$, line in the fourth quadrant), we obtain (since $\tilde v_{\rm max}>1>v$, but $v\tilde v_{\rm 
max}$ can be either positive or negative)
\be
\tilde v_{\rm max}\geq \frac{1}{v}\Rightarrow \tilde v^2_{\rm max}-1\leq \frac{1-v^2}{v^2}\Rightarrow
v^2_{\rm max}-1\leq \frac{1}{v^2}.
\ee

In the first case (when $1-v\tilde v_{\rm max}\leq 0$, and so $v^2_{\rm max}\leq 1+1/v^2$), we obtain the condition 
\be
(v_{\rm max}-v)^2\geq (1+(1-v^2(v^2_{\rm max}-1))(1-vv_{\rm max})^2.
\ee

We first note that at $v\rightarrow 1$, we obtain equality, so that seems encouraging. However, at $v=1-\epsilon$, we obtain 
\be
(v_{\rm max}-1)^2+2\epsilon(v_{\rm max}-1)\geq (v_{\rm max}-1)^2+2\epsilon(v_{\rm max}-1)
[(v_{\rm max}-1)^2(v_{\rm max}+1)-v_{\rm max}]\;,
\ee
which is not satisfied for any $v^2_{\rm max}\geq 1+1/v^2$. Since the velocity $v$ is an arbitrary parameter, as we 
can boost the embedding Minkowskian system of coordinates to any velocity, that is all we need. 

For completeness however, let us consider also the case of $v\ll 1$, which means that now 
\be
v^2_{\rm max}-1\geq \frac{1}{v^2}\gg 1\Rightarrow vv_{\rm max}\geq 1.
\ee

In this case, we see that the inequality is satisfied if $vv_{\rm max}-1\geq 1$, so if 
\be
\frac{\sqrt{1+v^2}}{v}\leq v_{\rm max} \leq \frac{2}{v}.
\ee

That means that if $v\ll 1$, we can extend the bound on $v_{\rm max}$ to be causal to $v_{\rm max}\leq 2/v$. 

What interests however is the bound obtained by minimizing over all possible $v$'s, and in that case, as we saw, we 
obtain (considering the most constraining case of $v\rightarrow 1$), 
\be
v^2_{\rm max}\leq 1+\frac{1}{v^2}\rightarrow2\Rightarrow v_{\rm max}\leq  \sqrt{2}.
\ee

This turns into a constraint on the $g_{0'0'}$ that can be created, 
\be
g_{0'0'}\leq 1.
\ee

It is not clear where would such a constraint come from. Perhaps it is a signal that we cannot construct a $g_{0'0'}>1$ 
with  any matter VEV. We leave this for further work.

\section{Fields, quantization and observations}

In this section we build up the non-Lorentz violating physical implications of the model, and explore whether there are any clear experimental signatures for 
the model. 

First, we consider the parameters of the model. Since the gravitational action is written in 5 dimensions, we have to reduce to 4 dimensions, on the compact
time direction. As we said, $R_0$ is the parameter that replaces the fixed ratio $dt'/dt=1$. Since the physical periodicity of $t'$ is $2\pi R_0\sqrt{g_{0'0'}}$, 
we can choose $R_0$ and $M_{\rm Pl,5}$ to be related as $2\pi R_0=R_{\rm Pl,5}=M_{\rm Pl,5}^{-1}$. 

From compactification of the 5 dimensional action on the second time, we obtain 
\be
2\pi R_0 \sqrt{g_{0'0'}}=R_{\rm Pl,5}\frac{M^2_{\rm Pl,4}}{M^2_{\rm Pl,5}}\;,
\ee
and from $2\pi R_0=R_{\rm Pl,5}$, we find
\be
g_{0'0'}=\left(\frac{M_{\rm Pl,4}}{M_{\rm Pl,5}}\right)^4.
\ee

Then it follows that we need $M_{\rm Pl,5}\gg M_{\rm Pl,4}$. For instance, with $M_{\rm Pl,5}=10^3 M_{\rm Pl,4}$, we obtain 
$g_{0'0'}\sim 10^{-12}$. 

Next, we turn to how to write fields and to quantize them in the presence of the second time. 

Unlike space, time {\em has} to flow, $dt'>0$, so for matter and its fields, we must have an $x_{0'}$ dependence. 
But then, if $x_{0'}$ is compact, it means that $e^{iE'x_{0'}\sqrt{g_{0'0'}}}$ must be single valued around the circle, 
which means that 
\be
E'=\frac{n}{R_0\sqrt{g_{0'0'}}}
\ee
so, if it is not zero, it is very large (at the Planck scale). Then, it can be the case that $E'=0$ for all physical processes we observe
today, in which case when quantizing, we just write wavefunctions
\be
e^{-iEt+i\vec{k}\cdot \vec{x}}\cdot e^{-iE' t' \sqrt{g_{0'0'}}}\rightarrow
e^{-iEt+i\vec{k}\cdot \vec{x}}\;,
\ee
so everything is standard, despite the fact that $dt'/dt=1$! That means that we don't have to change anything in the quantization of fields for 
energies below the Planck scale, which is what we would have wanted for consistency anyway.

A natural question that arises in the context of a quantum theory with a second time is the 
question of unitarity. In the case of fields, and of being in the vacuum, as we said, we have $n=0$, 
so fields are not excited in the second time, and nothing changes. But in the case of excited states with 
$n>0$, or of usual quantum mechanics of particles, we have a legitimate question. It is beyond the scope
of this paper to define a full quantum mechanical theory, but we just note that, in the special FLRW
frame, it is natural to construct a Schrodinger equation in terms of time evolution in the affine 
parameter (``proper time'' $\tau$ of eq. (\ref{propertime})), instead of the usual time $t$. 
In this case, unitarity works in the same way as for a single time $t$, and there is no contradiction.
The question of how to construct a fully covariant quantum theory, in particular for fields, is 
a hard one however. It is also less obvious what to do in the case of fields with $n>0$, but the 
quantum mechanical case suggests that, at least in the FLRW frame, viewing the field as a collection
of particles leads to no problems with unitarity, since the times $t$ and $t'$ flow together for 
these particles, and one
just has to be careful to define probabilities and their evolution 
with respect to the evolution in $\tau$, not $t$.

One assumption about our model was based on the quantization of matter fields: we assumed that we can vary $m$, as the VEV of some matter
fields, like some function of the Higgs scalar, or a fermion bilinear getting a VEV, or $\Tr[F_{\mu\nu}F^{\mu\nu}]$. Specifically, we want 
\be
m=[\langle \bar \psi \psi\rangle]^{1/3}\;\;\;{\rm or}\;\;\; \left(\Tr[F_{\mu\nu}F^{\mu\nu}]\right)^{1/4}\;\;\;{\rm or}\;\;\;
\left[|H|^2\right]^{1/2}.
\ee

{\bf Observations}

Besides directly activating the second time, i.e., increasing $g_{0'0'}$ as shown in the rest of the paper, the same lack of new phenomena below the 
Planck scale that we just mentioned also means that it is hard to pinpoint a smoking gun for the model. One could observe the presence of an infinitesimal 
$F_{0'123}$ perhaps, just that we chose it for the reason that it is consistent with the CMBR, so it would be hard to disentangle experimentally from it. 
We mentioned that $m$ must appear from some VEVs of matter fields, but normally it has to be infinitesimal too, so that is also difficult to observe. 

The one remaining possibility involves the field $\lambda$. If it is a Lagrange multiplier, there is nothing to observe, but if it is dynamical, 
there is a chance. The first observation is that $\d_{0'}\lambda=0$ if $E'=0$, even though in its kinetic term,
we have $(\d_{0'}\lambda)^2 g^{0'0'}$, with $g^{0'0'}\simeq 1/g_{0'0'}\gg 1$. So we cannot observe directly the second time through it either.

However, in this model, we have the quantum process $\lambda\rightarrow \phi\phi$, or rather, $\lambda\rightarrow 
A_{(3)0'12}A_{(3)0'12}$, and perhaps $\lambda $ can be identified with the Higgs. If we consider $\phi$ as the field, a scalar decaying in two other 
scalars is not much of a smoking gun, but as we saw, in the presence of gravity, the scalar and 3-form behave differently. If one could experimentally 
decide that the decay is really into two 3-forms, and not two scalars, then we would have a better test of the model.

\section{Conclusions}

In this paper I have shown that, if we consider the possibility of a small compact second time coordinate, under certain conditions we can have super-luminal 
propagation that is both causal, and respects general coordinate invariance. One needs a specific model, where a fundamental totally antisymmetric
3-form field $A_{(3)MNP}$ can get a nonzero field strength $F_{(4)0'123}$ via a coupling to the VEV $m$ of some matter fields, assumed to be possible 
to vary. In turn, $F_{(4)0'123}$, consistent with the FLRW Universe, can increase $g_{0'0'}$, increasing the physical (effective) radius of the second 
time. Motivating 4 physical principles (from the physics that we should observe from them) related to the presence of the second time, I derived from them
that we have an apparent Lorentz violation of $v^2_{\rm max}-1=g_{0'0'}/\gamma^2$, that is causal, 
at least for $v_{\rm max} \leq \sqrt{2}$, since $\gamma$ refers to the boost away from the 
FLRW cosmic time reference system. 

It should be noted that I have only proved that {\em if} we can construct (enhance) a macroscopic $g_{0'0'}$ around the (super-luminal) trajectory of a
macroscopic object ("spaceship"), then we obtain the causal apparent Lorentz violation $v^2_{\rm max}-1$ {\em from the point of view of the 
observer far away} (i.e., "on Earth"). The issue of how to actually construct this $g_{0'0'}$ (via increasing the VEV $m$ of matter fields) around the 
trajectory was not discussed. Therefore the present paper must be understood as a proof of principle of the possibility of causal super-luminal 
macroscopic propagation in the presence of the second time coordinate, and not as a way to actually construct such propagation.

\section*{Acknowledgements}
My work  is supported in part by CNPq grant 304006/2016-5 and FAPESP grant 2014/18634-9. I would also 
like to thank the ICTP-SAIFR for their support through FAPESP grant 2016/01343-7, and the University of Cape Town for hospitality during the period in which this 
work was being finalized.



\bibliography{Lorentz}
\bibliographystyle{utphys}

\end{document}